\documentclass[useAMS,usenatbib]{mn2e}
\usepackage{natbib}
\usepackage{amsmath}
\usepackage{amssymb}
\usepackage{graphicx}
\usepackage{color}
\usepackage{multirow}
\usepackage{color}
\usepackage{enumerate}
\usepackage[table]{xcolor}
\usepackage{float}
\usepackage{comment}
\usepackage{enumerate}
\usepackage{adjustbox}
\usepackage{subfigure}
\usepackage{epstopdf}
\begin{document}
\title[Probing the pulsar wind via timing of the other pulsar in a double pulsar binary]{Probing the properties of the pulsar wind via studying the dispersive effects in the pulses from the pulsar companion in a double neutron-star binary system}
\author[Yi \& Cheng]{Shu-Xu YI \thanks{yishuxu@hku.hk}, K. S. Cheng\\ Department of Physics, The University of Hong Kong, Pokfulam Road, Hong Kong}
\date{\today}
\maketitle
\begin{abstract}
The velocity and density distribution of $e^\pm$ in the pulsar wind are crucial distinction among magnetosphere models, and contains key parameters determining the high energy emission of pulsar binaries. In this work, a direct method is proposed, which might probe the properties of the wind from one pulsar in a double-pulsar binary. When the radio signals from the first-formed pulsar travel through the relativistic $e^\pm$ flow in the pulsar wind from the younger companion, the components of different radio frequencies will be dispersed. It will introduce an additional frequency-dependent time-of-arrival delay of pulses, which is function of the orbital phase. In this paper, we formulate the above-mentioned dispersive delay with the properties of the pulsar wind. As examples, we apply the formula to the double pulsar system PSR J0737-3039A/B and the pulsar-neutron star binary PSR B1913+16. For PSR J0737-3039A/B, the time delay in 300\,MHz is $\lesssim10\mu$s near the superior-conjunction, under the optimal pulsar wind parameters, which is $\sim$ half of the current timing accuracy. For PSR B1913+16, with the assumption that the neutron star companion has a typical spin down luminosity of $10^{33}$\,ergs/s, the time delay is as large as $10\sim20\mu$s in 300\,MHz. The best timing precision of this pulsar is $\sim5\mu$s in 1400\,MHz. Therefore, it is possible that we can find this signal in archival data. Otherwise, we can set an upper-limit on the spin down luminosity. Similar analysis can be apply to other eleven known pulsar-neutron star binaries.\\
\\
{\it Keywords}: binaries: general -- pulsars: general
\end{abstract}

\section{Introduction}
For its simplicity and intuitiveness, the vacuum magneto-dipole energy loss formula,
\begin{equation}
W^{(\rm{V})}_{\rm{tot}}=\frac{1}{6}\frac{B^2_0\Omega^4R^6}{c^3}\sin^2\chi,
\end{equation}
is still widely used as a rule of thumb to account for the braking of the spin-powered pulsars (see \citealt{2012hpa..book.....L} for instance). Although it has been known for a long time (see \citealt{1983ZhETF..85..401B} and references therein), that the magneto-dipole radiation should be fully screened by the magnetospheric plasma. More and more people believe that the pulsar wind takes away most of the rotational energy of the pulsar to infinity \citep{1969ApJ...158..727M,2013ApJ...768..144T,2014ApJ...788...16L,2016SCPMA..59a5752T}. The pulsar wind is composed of electromagnetic waves (EMW) and particles (mainly electrons/positions, $e^\pm$). The ratio between the energy fluxes of them is defined as the magnetization parameter $\sigma=W_{\rm{EM}}/W_{\rm{part}}$.

Studies with magnetohydrodynamic (MHD) theory demonstrated that particles can not be effectively accelerated in pulsar wind \citep{1975Ap&SS..32..375U,1996MNRAS.279.1168M,1998MNRAS.299..341B,1998ApJ...505..835C,1999MNRAS.305..211B,2001A&A...371.1155B,2001ApJ...562..494L,2002MNRAS.329L..34L}, therefore $\sigma\gg1$. However large kinetic energy of $e^\pm$ in pulsar wind is needed in modeling the observations in high energy. \cite{1984ApJ...283..694K} explained the luminosity of the Crab Nebula to be powered by the relativistic $e^\pm$ from the centre pulsar. A large kinetic energy in particles is required and thus $\sigma\ll1$ is implied. In $\gamma$-ray pulsar binaries, a large kinetic energy in particles is also crucial in explanation of the high energy emissions: In those models, the pulsar wind needs to collide with the stellar wind to form a termination shock front, which accounts for the KeV X-ray and TeV $\gamma$-ray emissions (e.g. in PSR B1259+63: \citealt{1999APh....10...31K,1999ApJ...521..718H,2005A&A...442....1A,2006MNRAS.367.1201C,2011MNRAS.416.1067K}; see also \citealt{2015CRPhy..16..661D} for a review). GeV emissions are modeled as outcomes of inverse Compton scattering by the cold $e^\pm$ in the pulsar wind \citep{2013A&A...557A.127D,2017ApJ...844..114Y}. The conflict between the needs of large kinetic energy of particles in the pulsar wind, and the difficulty in effective particle acceleration is known as the ``$\sigma$-problem" \citep{2016arXiv161003365B}.

To understand how the energy is dissipated from the Poynting flow to particles, i.e., to solve the $\sigma$-problem, has the merit for both the magnetosphere theories and for modeling the interaction between the pulsar and surrounding materials:

Since 1999 \citep{1999ApJ...511..351C,1999MNRAS.305..211B,1999A&A...349.1017B}, studies have been refining the understanding of the magnetosphere and making progress towards the solution of the problem (see for instances \citealt{2003ApJ...591..366K,2006ApJ...648L..51S,2014MNRAS.443.2197B}). A satisfactory theory of magnetosphere is essential for modeling pulsar braking, which is the main cause of pulsar timing noises \citep{2010MNRAS.402.1027H}. Reducing the timing noise is the major stream of efforts in pulsar timing, so as to unveil small signals such as gravitational waves \citep{2016ApJ...821...13A,2016MNRAS.455.1665B,2016SCPMA..59h..95Y,2016MNRAS.461.1317Z}. Observational constrains on the density and velocity distribution of $e^\pm$ in pulsar wind can test and select among these theories of magnetosphere;

On the other hand, as mentioned above, in models explaining the emission from pulsar nebulae and $\gamma$-ray pulsar binaries, the energy conversion process from EMW to particles is always considered phenomenally by $\sigma$ as a function of the distance to the pulsar (as in \citealt{2011MNRAS.416.1067K}):
\begin{equation}
\sigma=\sigma_{\rm{L}}\left(\frac{r}{r_{\rm{L}}}\right)^{-\alpha_{\sigma}},
\label{eqn:sigma}
\end{equation}
where $r_{\rm{L}}$ is the radius of the light cylinder, $\sigma_{\rm{L}}$ and $\alpha_{\sigma}$ are free parameters to be fitted to observations.

Therefore, an independent way to study the $\sigma-r$ relationship is wanted, which can help to rule out some of the proposed models, and reduce the degree of freedom in other models.
\begin{figure}
\centering
\includegraphics[width=8.5cm]{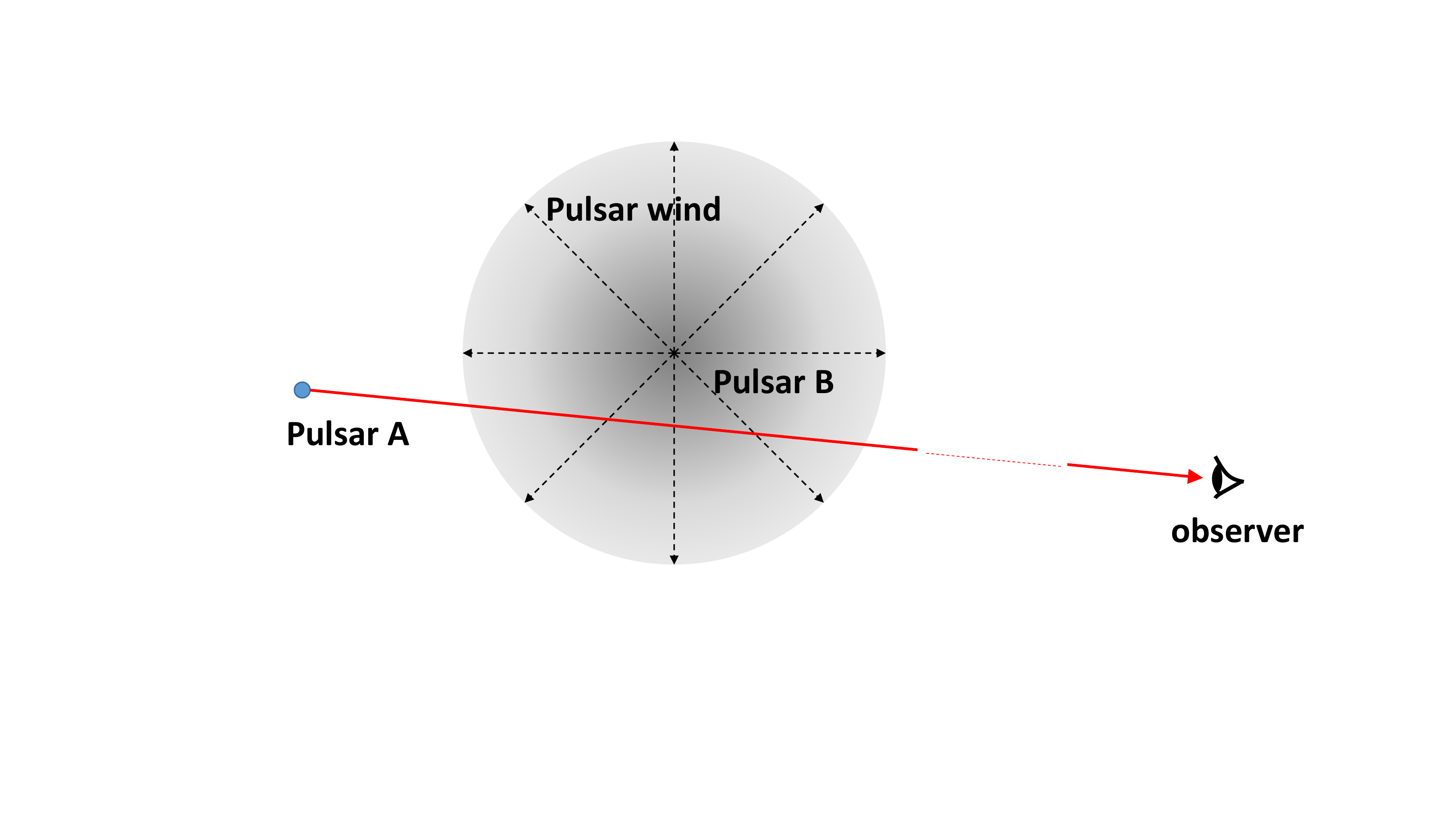}
\caption{The illustration of additional dispersion measure of one pulsar from the pulsar wind from the other pulsar in a double pulsar system}
\label{fig:illu}
\end{figure}

Double pulsar binary systems provide a possible method to study the above mentioned question. $e^\pm$ in the pulsar wind from one of the pulsar (pulsar B) act as dispersive medium, when the radio pulsations from the other pulsar in the binary (pulsar A) travel through it. The line of sight from the pulsar A to the Earth probes different depth in the pulsar wind at different orbital phases (see illustration in figure \ref{fig:illu}). Observation of the orbital phase-modulated dispersive effects in the signal of pulsar A can in principle serves as a way to study the density and velocity distribution of $e^\pm$ in the pulsar wind. When the bulk velocity of the $e^\pm$ is non-relativistic, the dispersion is determined solely by the density of $e^\pm$. When the medium becomes relativistic, the Lorentz factor of the bulk motion is also involved in the dispersion relationship. In section II, we study how the density and Lorentz factor distribution of $e^\pm$ in the pulsar wind, determines the additional dispersion measure (DM) of the pulses from pulsar A. We also show the time delay of the pulse arrival time (TOA) due to this additional DM. In section III, we apply our formula into two realistic binary systems: PSR J0737-3039A/B and J1915+1606. PSR J0737-3039A/B is the only known double pulsar binary. There are thirteen pulsar binary systems in which the companion is likely to be a neutron star (Double Neutron Stars or DNS, see the catalogue in \citealt{2017ApJ...835..185Y}). These invisible neutron star-companions can be pulsars whose radiating beams miss the line of sight. We choose the famous Hulse-Taylor binary (\citealt{1975ApJ...195L..51H}, also known as PSR J1915+1606 or B1913+16) as an example of the potential intrinsic double pulsar binaries. We show that pulsar timing observations which unveil such TOA variation signals can serve to determine the $\sigma-r$ relationship. In section IV, we discuss limits of practicability under the current radio telescope capabilities, and the further aspects of this proposed method. We conclude the paper in section V and discuss in section VI.

\section{Theory}
\begin{figure}
\centering
\includegraphics[width=7cm,trim={6cm, 3cm, 8cm 0cm},clip]{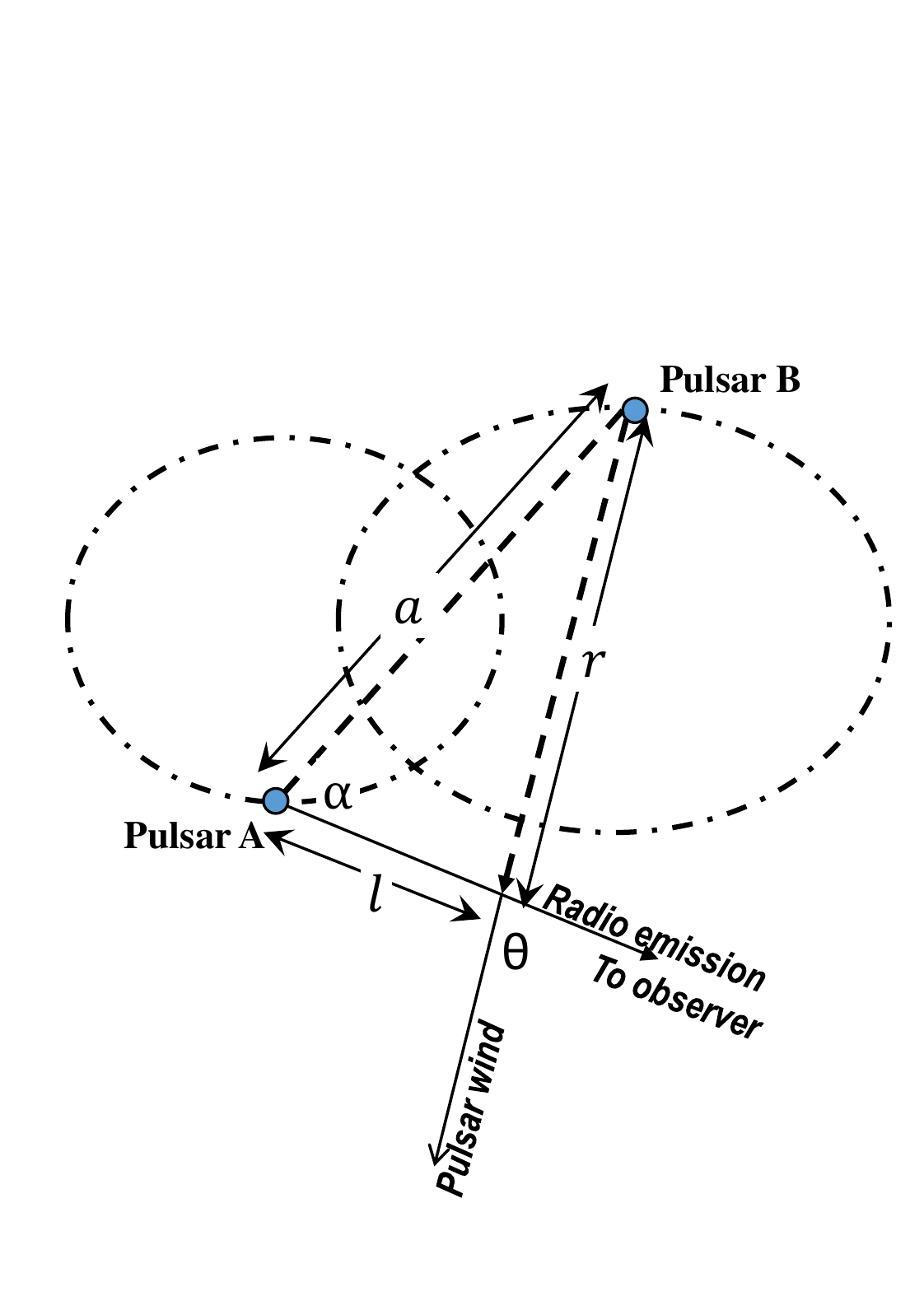}
\caption{The definitions of angles and distances. Pulsar B is the source of the pulsar wind, and pulsar A is the source of radio signals to be observed.}
\label{fig:1}
\end{figure}
In a double pulsar system, the radio emissions from pulsar A travel through the wind zone of pulsar B and are dispersed. In order to describe the dispersion process, we define the following quantities and angles: $a$ is the distance between pulsars A and B; $\alpha$ is the angle between the vector from pulsar A to B and the vector from pulsar A to the Earth; $l$ denotes the distance from the pulsar A to some point in the line of sight to the Earth; $\theta$ is the angle between the wind velocity and the propagating direction of the signals from A at that point (see figure 1 for illustration).

We derive the refractive index $n_{\nu}$ (defined as the group velocity of electromagnetic wave in the vacuum $c$ divided by that in the medium $c_{\rm{g}}$, $n_{\nu}\equiv c/c_{\rm{g}}$) of a stream of cold plasma with bulk velocity $\beta c$ as follows:

The Lorentz transformation of three dimensional velocity is:
\begin{equation}
u^\prime=\frac{\sqrt{u^2+\beta^2c^2-2\beta cu\cos\theta-\beta^2u^2\sin^2\theta}}{1-\beta u\cos\theta/c}.
\label{eqn:A1}
\end{equation}
In our case, the $u$ is specified to $c_{\rm{g}}$.
As definition, $c/c_{\rm{g}}^\prime=n^\prime_{\nu^\prime}$ is the refractive index in the stream co-moving frame (we will omit the prime mark over the subscript $\nu$ in the following text for pithiness), and $c/c_{\rm{g}}=n_\nu$ is the refractive index in the frame of the binary's barycentre.
Square equation (\ref{eqn:A1}), and note that:
$$1/n^\prime_\nu=\sqrt{1-\nu^2_{\rm{p}}/{\nu^\prime}^2},$$
($\nu_{\rm{p}}$ is the plasma frequency to be defined below), we have the equation about $n_\nu$:
\begin{equation}
\frac{n^2_\nu-1}{(n_\nu-\beta\cos\theta)^2}=\big(\frac{\nu_{\rm{p}}}{\nu}\big)^2\frac{1}{(1-\beta\cos\theta)^2}.
\end{equation}
Solve the above equation and constrain the $\nu$ of interest that $\nu\gg\nu_{\rm{p}}$. To the lowest order of $\nu_{\rm{p}}/\nu$, the solution is:
\begin{equation}
n(\nu)=1+\frac{1}{2}\frac{1}{1-\beta\cos\theta}\big(\frac{\nu_{\rm{p}}}{\nu}\big)^2,
\label{eqn:1}
\end{equation}
where $\nu$ is the frequency of the electromagnetic wave, and the plasma frequency is:
\begin{equation}
\nu_{\rm{p}}=\sqrt{\frac{e^2n^\prime_{\rm{e}}}{\pi m_{\rm{e}}}}\approx8.5\,\text{kHz}\big(\frac{n^\prime_{\rm{e}}}{\text{cm}^3}\big)^{1/2},
\label{eqn:nu}
\end{equation}
where $n^\prime_{\rm{e}}$ is the electron number density in the stream co-moving frame.

The above derivations ignored the gravitational redshift of the gravity of pulsar B. The gravitational redshift $z(r)\sim r_{\rm{s}}/(2r)$, which is $\ll1$ in the region where we are interested. $r_{\rm{s}}\equiv2GM_{\rm{B}}/c^2$ is the Schwarzschild radius of pulsar B.

For an isotropic pulsar wind, the electron number density in the barycentric frame at a distance $r$ is:
\begin{equation}
n_{\rm{e}}(r)=\frac{L_{\rm{sd}}}{4\pi\beta c^3r^2m_{\rm{e}}(1+\sigma)\gamma},
\label{eqn:3}
\end{equation}
where $L_{\rm{sd}}$ is the spin down power, and $\sigma$ is the magnetization parameter, which is defined as the ratio between the energy fluxes in the Poynting flow and that in the particles, $\gamma$ is the Lorentz factor of the bulk velocity of the wind. The energy in the pulsar wind transfers from the Poynting flow to $e^\pm$ outward gradually. As $r\rightarrow\infty$, all the energy in the pulsar wind is transferred to electrons, thus $\sigma\rightarrow0$ and $\gamma\rightarrow\gamma_\infty$. Due to the conservation of electrons in the region far outside the magnetosphere, $(1+\sigma)\gamma\beta$ remains constant and equal to $\gamma_\infty$ (where $\beta\rightarrow1$).
\begin{equation}
n_{\rm{e}}^\prime=n_{\rm{e}}/\gamma.
\label{eqn:nprime}
\end{equation}

The $\cos\theta$ in equation (\ref{eqn:1}) is given by the cosine theorem:
\begin{equation}
\cos\theta=\frac{l-a\cos\alpha}{\sqrt{a^2+l^2-2al\cos\alpha}},
\end{equation}
where the notations are defined in above paragraph and in the illustrating figure 1, and:
\begin{equation}
\cos\alpha=-\sin\theta_\oplus\cos(\phi_\oplus-\varphi),
\end{equation}
where $\varphi$ and $\phi_\oplus$ are the true anomaly and the azimuthal angle of the observer from the periastron of pulsar B respectively, and $\theta_{\oplus}$ is the polar angle of the observer from the normal vector of the orbital plane. In the above derivation of geometry relationship, we assume the trajectories of light are straight lines. It is because we constrain our treatment where the pulse from pulsar A will not be eclipsed by the magnetosphere of pulsar B. Therefore the Einstein angle of light deflection $\theta_{\rm{Einstein}}=2r_{\rm{s}}/r\ll1$, where $r_{\rm{s}}\sim10^5$\,cm is the Schwarzschild radius of the pulsar B, and $r$ is the impact distance of the light rays, which should larger than the radius of the light cylinder ($10^8$\,cm).

The time lag between a signal of infinity frequency and $\nu$ is:
\begin{equation}
\delta t=\int^\infty_0\frac{n(\nu)dl}{c\sqrt{1-r_{\rm{s}}/r}}-\int^\infty_0\frac{dl}{c\sqrt{1-r_{\rm{s}}/r}},
\label{eqn:12}
\end{equation}
where the factor $\sqrt{1-r_{\rm{s}}/r}$ accounts for the time dilation of general relativity, $dl$ is the distance element away from the pulsar. With the equation (\ref{eqn:1}), the extra DM due to the pulsar wind is
\begin{equation}
\delta\text{DM}=\int^\infty_0\frac{1}{\sqrt{1-r_{\rm{s}}/r}}\frac{n^\prime_{\rm{e}}}{(1-\beta\cos\theta)}dl.
\end{equation}
The above integration starts from the pulsar and goes to infinity where the density of electrons in the pulsar wind is zero. 

\section{Applications to existing double pulsar systems}
\subsection{PSR J0737-3039A/B}
PSR J0737-3039A/B is the only known system that both of its companions are radio active pulsars. The parameters of this system are \citep{2008ARA&A..46..541K}:
\begin{itemize}
\item{Orbital inclination angle: $88.69^\circ$}
\item{Longitude of periastron: $87.0331^\circ$}
\item{Eccentricity, $e=0.087775$}
\item{Projected semimajor axis, $x=(a/c)\sin i$: \\pulsar A: 1.415032\,s; pulsar B: 1.5161\,s}
\item{Spin down luminosity of pulsar B, $L_{\rm{sd,B}}=1.7\times10^{30}$\,erg/s.}
\end{itemize}
The bulk Lorentz factor of the pulsar wind as function of the distance is:
\begin{equation}
\gamma(r)=\sqrt{1+\left(\frac{\gamma_\infty}{1+\sigma(r)}\right)^2},
\end{equation}
where $\sigma(r)$ is described with equation (\ref{eqn:sigma}), and
\begin{equation}
\sigma_{\rm{L}}=\frac{B^2_{\rm{L}}/8\pi}{2\dot{N}_{\rm{e^\pm}}m_ec/r^2_{\rm{L}}},
\end{equation}
where $B_{\rm{L}}$ is the magnetic field at the light cylinder, $r_{\rm{L}}$ is the radius of the light cylinder, $\dot{N}_{\rm{e^\pm}}=N_{\rm{m}}\dot{N}_{\rm{GJ}}$, $N_{\rm{m}}$ is the multiplicity of $e^\pm$ and $\dot{N}_{\rm{GJ}}$ is the Goldreich-Julian particle flow rate at the light cylinder. \cite{2011MNRAS.416.1067K} evaluated the $\sigma_{\rm{L}}$ of the Crab pulsar to be $\sim10^5$.

\begin{figure}
\centering
\includegraphics[width=7cm]{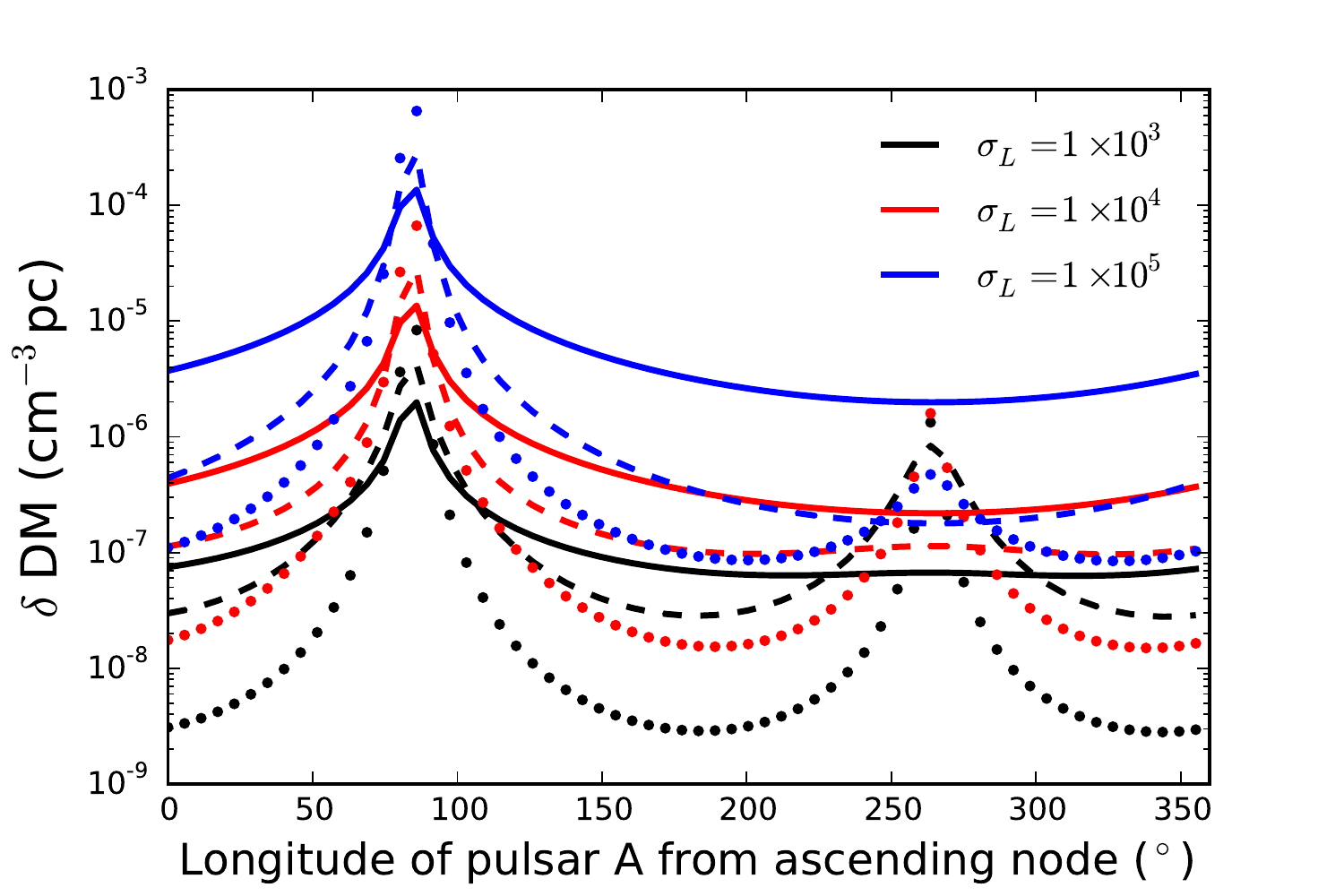}
\caption{The additional DM in PSR J0737-3039A, under the assumptions of different $\sigma_L$ and $\alpha_\sigma$. The black, red and blue line colors correspond to $\sigma_L=1\times10^3$, $1\times10^4$ and $1\times10^5$ respectively; The solid, dashed and dotted line styles correspond to $\alpha_\sigma=0$, $1$, $2$. For all curves, $\gamma_\infty=10^3$ is adopted.}
\label{fig:2}
\end{figure}

With above parameters, we plot the $\delta DM$ as function of the orbital phase (longitude from the ascending node) in figure \ref{fig:2}, under different $\sigma_L$ and $\alpha_\sigma$. The orbital phases when pulsar A is eclipsed by the magnetosphere of pulsar B should be excluded from above figures \ref{fig:1} and \ref{fig:2}. 

\begin{figure}
\centering
\includegraphics[width=7cm]{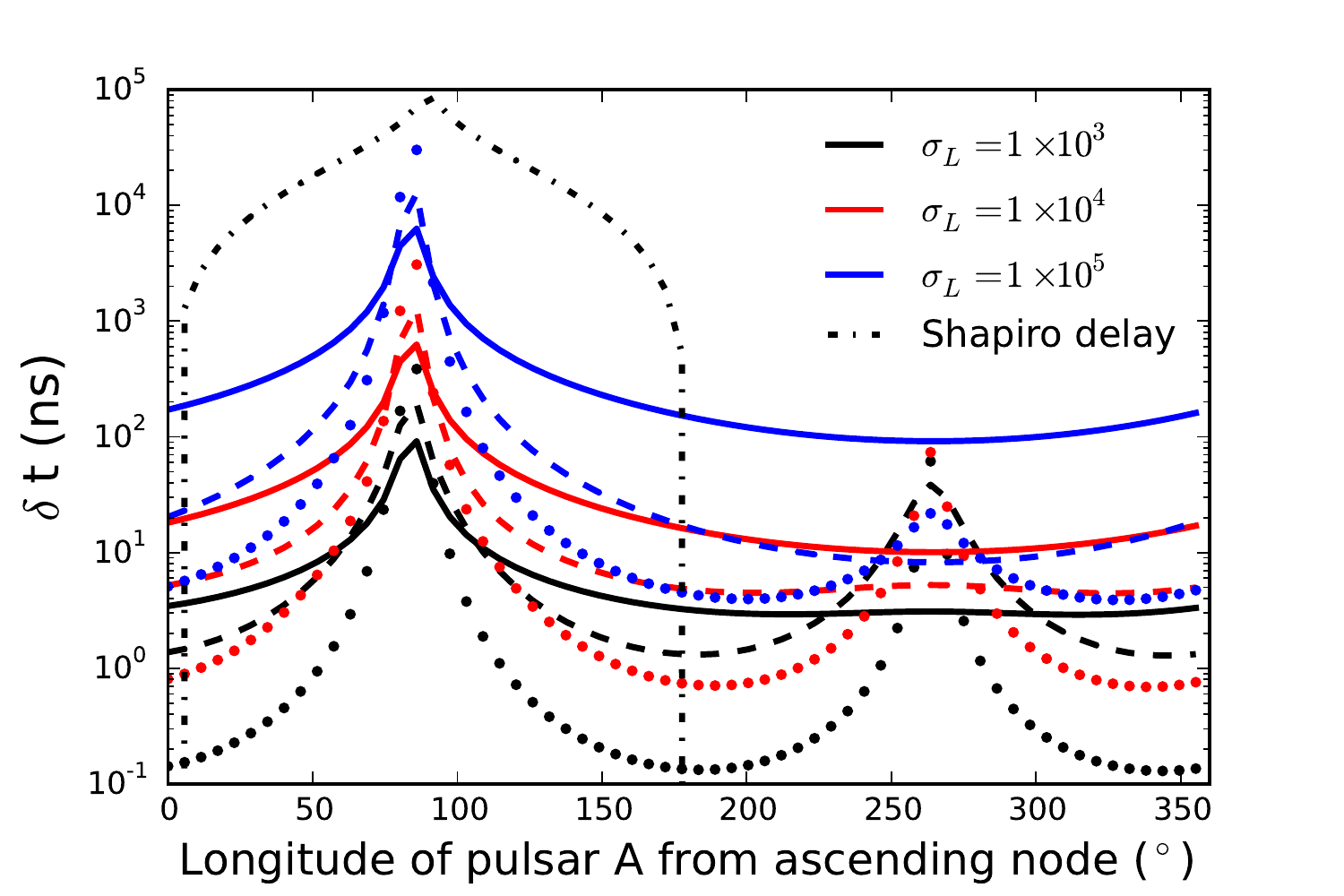}
\caption{Time delay due to the additional DM in PSR J0737-3039A/B in the observing frequency of 300\,MHz, compared with Shapiro delay (dash-dotted curve). The black, red and blue line colors correspond to $\sigma_L=1\times10^3$, $1\times10^4$ and $1\times10^5$ respectively; The solid, dashed and dotted line styles correspond to $\alpha_\sigma=0$, $1$ and $2$ respectively. For all curves, $\gamma_\infty=10^3$ is adopted.}
\label{fig:3}
\end{figure}

The additional DM causes delay of TOA of the pulses from pulsar A:
\begin{equation}
\Delta t=4.15\times10^6\,\text{ms}\times f^{-2}\times\delta DM,
\end{equation}
where $f$ is the radio frequency in which pulsar A is observed. Figure \ref{fig:3} plots the $\Delta t$ corresponding to the $\delta DM$, when the observing frequency is 300\,MHz.

When producing both of the figures, $\gamma_\infty=10^3$ is adopted. As are shown in figures \ref{fig:2} and \ref{fig:3}, $\delta DM$ and $\Delta t$ increase toward the superior-conjunction of pulsar A (when pulsar A is behind the pulsar B). It is because that the path of signals from pulsar A is longer, with denser wind region in between the line of sight. In the cases of $\sigma_{\rm{L}}=10^3$,  $\delta DM$ and $\Delta t$ also increase toward the inferior-conjunction of pulsar A (when pulsar A is in front of the pulsar B). It is because that with smaller $\sigma_{\rm{L}}$, the $e^\pm$ flow in the pulsar wind quickly becomes relativistic ($\beta\sim1$). As a result, in the inferior-conjunction where $\cos\theta\rightarrow1$, the factor $1/(1-\beta\cos\theta)$ increases significantly as $\beta\rightarrow1$.
\subsection{PSR J1915+1606 (B1913+16) as an example of DNS}
PSR J1915+1606 (B1913+16), also known as the Hulse-Taylor binary, consists of two neutron stars. Only one of the neutron star is detected as radio pulsar. The parameters of this system are \citep{2016ApJ...829...55W}:
\begin{itemize}
\item{Orbital inclination angle: $42.84^\circ$}
\item{Longitude of periastron: $292.54450^\circ$}
\item{Eccentricity, $e=0.6171340$}
\item{Projected semimajor axis, $x=(a/c)\sin i=2.341776$\,s}
\item{Mass ratio between the pulsar and the companion neutron star: 1.0345}
\end{itemize}

We refer to the quite neutron star as the ``pulsar B" although it is not detected as a pulsar. Since spin period and spin frequency derivative is not known, $r_{\rm{L}}$ and $L_{\rm{sd}}$ of it needs to be assumed. We assume $r_{\rm{L}}$ to be the same as PSR J0737-3039b, and $L_{\rm{sd}}=10^{33}$\,ergs/s as a typical pulsar's spin down luminosity. Besides, $\gamma_\infty=10^3$ is adopted.

\begin{figure}
\centering
\includegraphics[width=7cm]{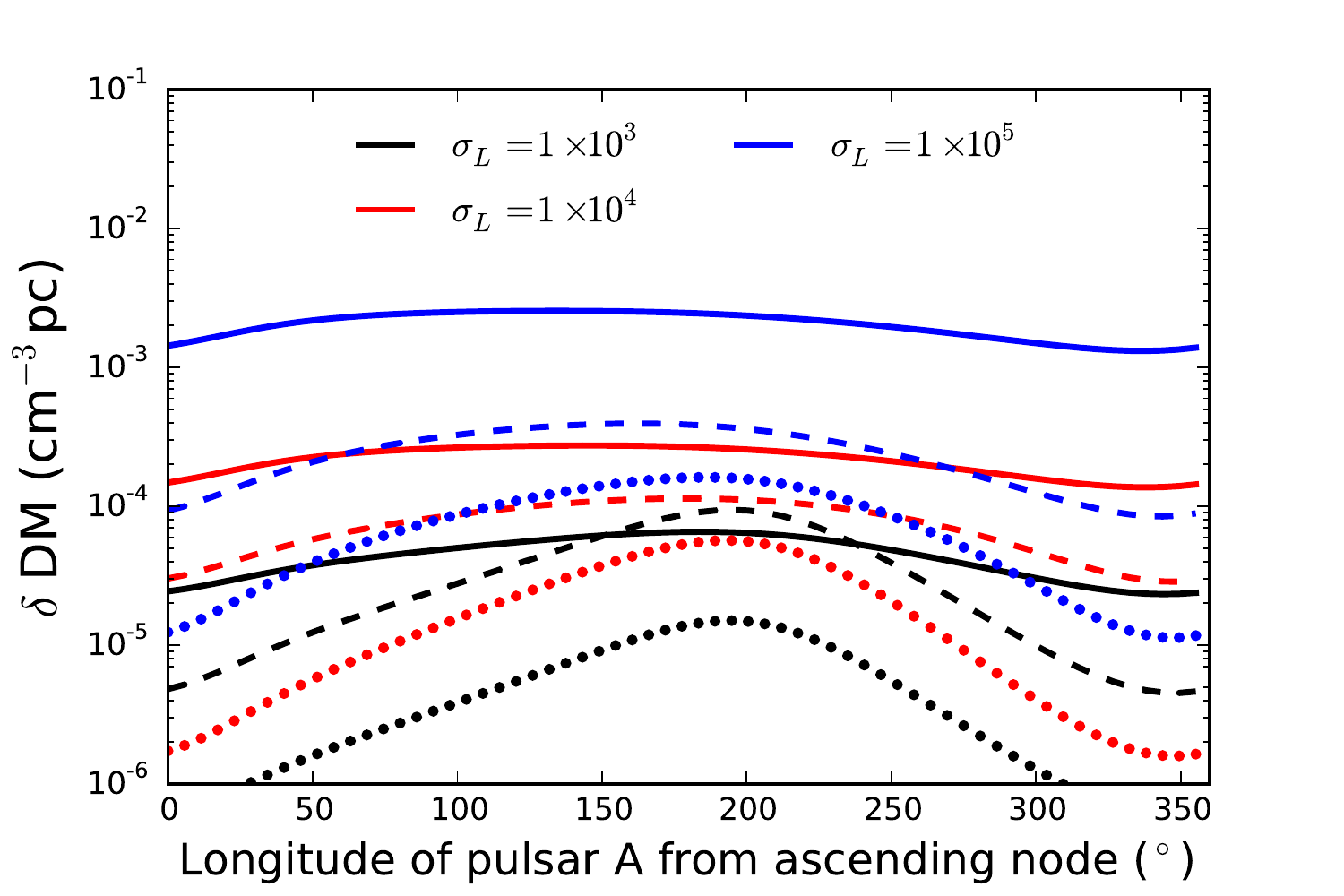}
\caption{The additional DM of PSR J1915+1606, under the assumptions of different $\sigma_L$ and $\alpha_\sigma$. The black, red and blue line colors correspond to $\sigma_L=1\times10^3$, $1\times10^4$ and $1\times10^5$ respectively; The solid, dashed and dotted line styles correspond to $\alpha_\sigma=0$, $1$, $2$. For all curves, $\gamma_\infty=10^3$ is adopted.}\label{fig:5}
\end{figure}

\begin{figure}
\centering
\includegraphics[width=7cm]{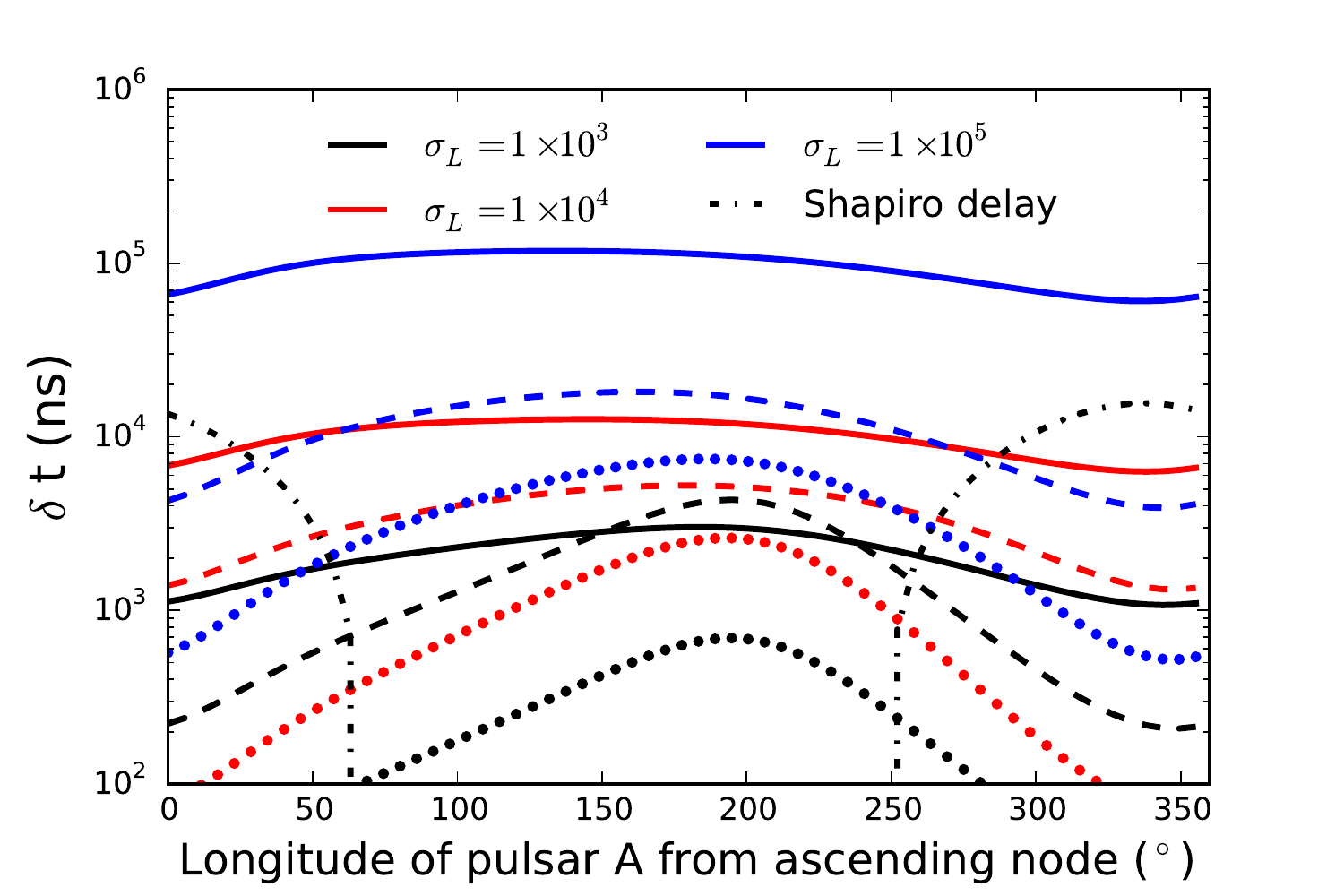}
\caption{Time delay due to the additional DM in PSR J1915+1606 in the observing frequency of 300\,MHz, compared with Shapiro delay (dash-dotted curve). The black, red and blue line colors correspond to $\sigma_L=1\times10^3$, $1\times10^4$ and $1\times10^5$ respectively; The solid, dashed and dotted line styles correspond to $\alpha_\sigma=0$, $1$ and $2$ respectively. For all curves, $\gamma_\infty=10^3$ is adopted.}\label{fig:4}
\end{figure}

Similar with calculations in the case of PSR J0737-3039A/B, we plot $\delta DM$ and $\Delta t$ in figure (\ref{fig:5},\ref{fig:4}) with different $\sigma_L$ and $\alpha_\sigma$.
Figure \ref{fig:4} plots the corresponding time delay due to the additional DM in 300\,MHz (see details in the caption of the figure).
In contract with PSR J0737-3039A/B, the mild inclination angle (42.84$^\circ$) and large eccentricity ($e=0.6171340$) make the peaks of $\delta DM$ and $\Delta t$ locate differently from that of the Shapiro delay, i.e. the superior-conjunction.
\section{Feasibility of archival, current and further observations}
The best timing work of PSR J0737-3039A/B so far is \cite{2006Sci...314...97K}. In that study, the pulsar was observed at six frequency bands of three large radio telescopes (64-m Parkes radio telescope, 76-m Lovell radio telescope and 100-m Green Bank Telescope (GBT)). The observing frequencies range from 340\,MHz to 3030\,MHz. The best timing precision of the pulsar A was obtained at 820\,MHz of GBT, with typical TOA uncertainties of 18\,$\mu$s with a 30-s integration.

From the analysis in the above section (see figure \ref{fig:3}), in an optimistic case when $\sigma_{\rm{L}}=10^5$ and $\gamma_\infty=10^3$, the predicted time delay is $\sim$10\,$\mu$s at 300\,MHz. At least 4 times longer integration of GBT data will be needed to improve the timing precision two times better than the current $\sim18\,\mu$s. If the pulse profiles broadening towards low frequencies (either intrinsic or due to ISM scattering) is taking into consideration (see figure S.1 in \citealt{2006Sci...314...97K}), the integration should be longer up to $\sim3$ minutes. If we want to observe the signal as function of orbital phase, the timing precision needs to be improved 20 times than current value. In this case, the integration needs to be as long as 5 hours. Since the signal is orbital phase-dependent, we should not add up the data in different phases. Instead, we average the data in the same phase from different orbits. Observations of $\sim$20 orbits will be needed to obtain a resolution of 0.1 phase. In order to do that, a precious modeling of the peculiar evolution over orbits due to relativistic effects \citep{2014IJMPD..2330004K} is crucial.

Larger radio telescopes, e.g. FAST \citep{2016RaSc...51.1060L}, SKA \citep{2017ARep...61..288G}, giving better timing precision and/or lower observing frequencies, e.g. LOFAR \citep{2011AIPC.1357..325S}, are helpful towards resolving the signals. We will study the further observational aspects using simulated observations in following papers.

The spin down luminosity of PSR J0737-3039A is $\sim3000$ times larger than the pulsar B. Thus we expect the time delay of signals from pulsar B due to the pulsar wind of pulsar A could be $\sim3000$ times larger than that around, in the same observing frequencies. However, the emission of pulsar B is strongly influenced by the wind from pulsar A \citep{2004ApJ...614L..53Z}, thus the pulse profiles are orbital phase-dependent \citep{2005ApJ...624L.113B}. Furthermore, there is a significant spin precession by $5.1^\circ$/yr \cite{2005ApJ...624L.113B} as a result of general relativity coupling of spin and the total angular momentum \cite{1974CRASM.279..971B}. Such orbital and secular evolution of the pulse profiles make the timing precision of pulsar B much less than that of pulsar A (in \citealt{2006Sci...314...97K}, the RMS timing residual of the pulsar B with 300-s integration about 400 times larger than that of pulsar A).

The difficulties of observation in system PSR J0737-3039A/B is mainly due to the low spin down luminosity of the pulsar B. It is common to have $L_{\rm{sd}}=10^{33}\sim10^{35}$\,ergs/s in pulsars. If in a DNS binary $L_{\rm{sd}}$ of the neutron star companion is $10^{33}\sim10^{35}$\,ergs/s, the proposed features might have already been recorded in the archival data. As an example, the best up to date timing observation of PSR B1913+16 is done by \cite{2016ApJ...829...55W}. In their work, thirty-one years of data from Arecibo Observatory at 1400\,MHz were used. With 5-minute integration, the TOA uncertainties were obtained as small as $\sim5\,\mu$s. As shown in the above section (see figure \ref{fig:4}), the predicted time delay may ready to be seen ($10\sim20\mu$s) in an archival data in 300\,MHz with the same timing precision, or at least a useful limit of the properties of the pulsar wind of the invisible neutron star companion can be set from the data. The observation of the dispersive effects from the neutron star companions provides an unique way to determine the pulsar nature of the invisible neutron stars, which can not be probe otherwise.
\section{Summary}
We studied the dispersive effects of the signal of a pulsar, arising from the pulsar wind of the other pulsar in a double-pulsar system. The resulted additional dispersion measure (DM) is formulated related to the properties of the pulsar wind. We applied the formula to the only known double-pulsar binary PSR J0737-3039A/B, and the Hulse-Taylor binary as an example of potential intrinsic double-pulsar binaries.
The conclusions of this work are listed below:
\begin{enumerate}[1.]
\item{Additional DM and the resulted time delay as functions of the orbital phase are able to reflect properties of the pulsar wind. These properties are: the spin down luminosity $L_{\rm{sd}}$, the magnetization parameter $\sigma$ as function of distance to the wind source, the asymptotic Lorentz factor of the $e^\pm$ in the wind $\gamma_\infty$. See figures \ref{fig:2},\ref{fig:3},\ref{fig:5},\ref{fig:4}}.
\item{For PSR J0737-3039A/B, the time delay in 300\,MHz is $\lesssim10\mu$s near the superior-conjunction. The time delay is inversely proportional to the square of the observing frequencies. The current best timing accuracy of J0737-3039A is $\sim$ two times larger than wanted. Therefore longer integration of data, further observations with larger telescopes and lower observing frequencies are needed. }
\item{With the assumption that the neutron star companion of PSR B1913+16 has a typical spin down luminosity of $10^{33}$\,ergs/s, the time delay is as large as $10\sim20\mu$s in 300\,MHz. The best timing precision of this pulsar is $\sim5\mu$s in 1400\,MHz. Therefore it is possible that we can find this signal in archival data. Otherwise, we can set an upper-limit on the spin down luminosity to $\sim10^{33}$\,ergs/s, under certain assumptions of the pulsar wind properties.}
\end{enumerate}
\section{Discussion}
The magnetosphere of PSR J0737-3039B is thought to be distorted by the wind from pulsar A \citep{2004MNRAS.353.1095L}. As a result, the wind from pulsar B should be anisotropic rather than what is assumed in this paper. However, we expected the wind zone of pulsar B is less deformed, since the pulsar wind is lepton dominated, and the interactions among leptons are weak. Besides, the spherically symmetric treatment above can serve as the zeroth-order approximation, before we can accurately model the directional dependence of the wind of pulsar B. 

As can be seen from equations (\ref{eqn:3}, \ref{eqn:nprime} and \ref{eqn:12}), the amplitude of the proposed signal is proportional to $L_{\rm{sd}}/\gamma_\infty^2$. Therefore, if no such signal is seen, the upper-limit will be actually set on the combination $L_{\rm{sd}}/\gamma_\infty^2$. The value of $\gamma_\infty$ varies from $10^3$ to $10^6$ as fitted values in different models. $\gamma_\infty=10^3$ is adopted throughout the calculations in this paper, as an optimized condition for the maximum TOA variation signals given $L_{\rm{sd}}$. Larger $\gamma_\infty$ will make the signal less likely to be seen. 
\section*{Acknowledgement}
We thank Prof. Kramer, Dr. Janssen and Prof. Lorimer for their instructive discussions on this idea. Prof. McLaughlin and Dr. Pol gave a lot of helpful suggestions on improving the manuscript. This work is partially supported by a GRF grant under 17302315.



\begin{thebibliography}{99}
\bibitem[Aharonian et al.(2005)]{2005A&A...442....1A} Aharonian, F., Akhperjanian, A.~G., Aye, K.-M., et al.\ 2005, A\&A, 442, 1
\bibitem[Aragona et al.(2010)]{2010ApJ...724..306A} Aragona, C., McSwain, M.~V., \& De Becker, M.\ 2010, ApJ, 724, 306
\bibitem[Arzoumanian et al.(2016)]{2016ApJ...821...13A} Arzoumanian, Z., Brazier, A., Burke-Spolaor, S., et al.\ 2016, ApJ, 821, 13
\bibitem[Babak et al.(2016)]{2016MNRAS.455.1665B} Babak, S., Petiteau, A., Sesana, A., et al.\ 2016, MNRAS, 455, 1665
\bibitem[Beskin et al.(1983)]{1983ZhETF..85..401B} Beskin, V.~S., Gurevich, A.~V., \& Istomin, I.~N.\ 1983, Zhurnal Eksperimentalnoi i Teoreticheskoi Fiziki, 85, 401
\bibitem[Beskin et al.(1998)]{1998MNRAS.299..341B} Beskin, V.~S., Kuznetsova, I.~V., \& Rafikov, R.~R.\ 1998, MNRAS, 299, 341
\bibitem[Beskin(2016)]{2016arXiv161003365B} Beskin, V.~S.\ 2016, arXiv:1610.03365
\bibitem[Bogovalov \& Tsinganos(1999)]{1999MNRAS.305..211B} Bogovalov, S., \& Tsinganos, K.\ 1999, MNRAS, 305, 211
\bibitem[Bogovalov(1999)]{1999A&A...349.1017B} Bogovalov, S.~V.\ 1999, A\&A, 349, 1017
\bibitem[Bogovalov(2001)]{2001A&A...371.1155B} Bogovalov, S.~V.\ 2001, A\&A, 371, 1155
\bibitem[Bogovalov(2014)]{2014MNRAS.443.2197B} Bogovalov, S.~V.\ 2014, MNRAS, 443, 2197
\bibitem[Burgay et al.(2005)]{2005ApJ...624L.113B} Burgay, M., Possenti, A., Manchester, R.~N., et al.\ 2005, ApJL, 624, L113
\bibitem[Casares et al.(2012)]{2012MNRAS.421.1103C} Casares, J., Rib{\'o}, M., Ribas, I., et al.\ 2012, MNRAS, 421, 1103
\bibitem[Chernyakova et al.(2006)]{2006MNRAS.367.1201C} Chernyakova, M., Neronov, A., Lutovinov, A., Rodriguez, J., \& Johnston, S.\ 2006, MNRAS, 367, 1201
\bibitem[Chiueh et al.(1998)]{1998ApJ...505..835C} Chiueh, T., Li, Z.-Y., \& Begelman, M.~C.\ 1998, ApJ, 505, 835
\bibitem[Contopoulos et al.(1999)]{1999ApJ...511..351C} Contopoulos, I., Kazanas, D., \& Fendt, C.\ 1999, ApJ, 511, 351
\bibitem[Damour \& Ruffini(1974)]{1974CRASM.279..971B} Damour, T., \& Ruffini, R.\ 1974, Academie des Sciences Paris Comptes Rendus Serie Sciences Mathematiques, 279, 971
\bibitem[Dubus \& Cerutti(2013)]{2013A&A...557A.127D} Dubus, G., \& Cerutti, B.\ 2013, A\&A, 557, A127
\bibitem[Dubus(2015)]{2015CRPhy..16..661D} Dubus, G.\ 2015, Comptes Rendus Physique, 16, 661
\bibitem[Grainge et al.(2017)]{2017ARep...61..288G} Grainge, K., Alachkar, B., Amy, S., et al.\ 2017, Astronomy Reports, 61, 288
\bibitem[Hirayama et al.(1999)]{1999ApJ...521..718H} Hirayama, M., Cominsky, L.~R., Kaspi, V.~M., et al.\ 1999, ApJ, 521, 718
\bibitem[Hobbs et al.(2010)]{2010MNRAS.402.1027H} Hobbs, G., Lyne, A.~G., \& Kramer, M.\ 2010, MNRAS, 402, 1027
\bibitem[Hulse \& Taylor(1975)]{1975ApJ...195L..51H} Hulse, R.~A., \& Taylor, J.~H.\ 1975, ApJL, 195, L51
\bibitem[Kennel \& Coroniti(1984)]{1984ApJ...283..694K} Kennel, C.~F., \& Coroniti, F.~V.\ 1984, ApJ, 283, 694
\bibitem[Kirk et al.(1999)]{1999APh....10...31K} Kirk, J.~G., Ball, L., \& Skj{\ae}raasen, O.\ 1999, Astroparticle Physics, 10, 31
\bibitem[Kirk \& Skj{\ae}raasen(2003)]{2003ApJ...591..366K} Kirk, J.~G., \& Skj{\ae}raasen, O.\ 2003, ApJ, 591, 366
\bibitem[Kramer et al.(2006)]{2006Sci...314...97K} Kramer, M., Stairs, I.~H., Manchester, R.~N., et al.\ 2006, Science, 314, 97
\bibitem[Kramer(2014)]{2014IJMPD..2330004K} Kramer, M.\ 2014, International Journal of Modern Physics D, 23, 1430004
\bibitem[Kong et al.(2011)]{2011MNRAS.416.1067K} Kong, S.~W., Yu, Y.~W., Huang, Y.~F., \& Cheng, K.~S.\ 2011, MNRAS, 416, 1067
\bibitem[Kramer \& Stairs(2008)]{2008ARA&A..46..541K} Kramer, M., \& Stairs, I.~H.\ 2008, Annu. Rev. Astron. Astrophys., 46, 541
\bibitem[Li et al.(2014)]{2014ApJ...788...16L} Li, L., Tong, H., Yan, W.~M., et al.\ 2014, ApJ, 788, 16
\bibitem[Li \& Pan(2016)]{2016RaSc...51.1060L} Li, D., \& Pan, Z.\ 2016, Radio Science, 51, 1060
\bibitem[Lorimer \& Kramer(2012)]{2012hpa..book.....L} Lorimer, D.~R., \& Kramer, M.\ 2012, Handbook of Pulsar Astronomy, by D.~R.~Lorimer , M.~Kramer, Cambridge, UK: Cambridge University Press, 2012,
\bibitem[Lyubarsky \& Eichler(2001)]{2001ApJ...562..494L} Lyubarsky, Y., \& Eichler, D.\ 2001, ApJ, 562, 494
\bibitem[Lyubarsky(2002)]{2002MNRAS.329L..34L} Lyubarsky, Y.~E.\ 2002, MNRAS, 329, L34
\bibitem[Lyutikov(2004)]{2004MNRAS.353.1095L} Lyutikov, M.\ 2004, MNRAS, 353, 1095
\bibitem[Melatos \& Melrose(1996)]{1996MNRAS.279.1168M} Melatos, A., \& Melrose, D.~B.\ 1996, MNRAS, 279, 1168
\bibitem[Michel(1969)]{1969ApJ...158..727M} Michel, F.~C.\ 1969, ApJ, 158, 727
\bibitem[Spitkovsky(2006)]{2006ApJ...648L..51S} Spitkovsky, A.\ 2006, ApJL, 648, L51
\bibitem[Stappers et al.(2011)]{2011AIPC.1357..325S} Stappers, B., Hessels, J., Alexov, A., et al.\ 2011, American Institute of Physics Conference Series, 1357, 325
\bibitem[Tong et al.(2013)]{2013ApJ...768..144T} Tong, H., Xu, R.~X., Song, L.~M., \& Qiao, G.~J.\ 2013, ApJ, 768, 144
\bibitem[Tong(2016)]{2016SCPMA..59a5752T} Tong, H.\ 2016, Science China Physics, Mechanics, and Astronomy, 59, 5752
\bibitem[Usov(1975)]{1975Ap&SS..32..375U} Usov, V.~V.\ 1975, AP\&SS, 32, 375
\bibitem[Waters et al.(1988)]{1988A&A...198..200W} Waters, L.~B.~F.~M., van den Heuvel, E.~P.~J., Taylor, A.~R., Habets, G.~M.~H.~J., \& Persi, P.\ 1988, A\&A, 198, 200
\bibitem[Weisberg \& Huang(2016)]{2016ApJ...829...55W} Weisberg, J.~M., \& Huang, Y.\ 2016, ApJ, 829, 55
\bibitem[Yang et al.(2017)]{2017ApJ...835..185Y} Yang, Y.-Y., Zhang, C.-M., Li, D., et al.\ 2017, ApJ, 835, 185
\bibitem[Yi \& Zhang(2016)]{2016SCPMA..59h..95Y} Yi, S.-X., \& Zhang, S.-N.\ 2016, Science China Physics, Mechanics, and Astronomy, 59, 95
\bibitem[Yi \& Cheng(2017)]{2017ApJ...844..114Y} Yi, S.-X., \& Cheng, K.~S.\ 2017, ApJ, 844, 114
\bibitem[Zhang \& Loeb(2004)]{2004ApJ...614L..53Z} Zhang, B., \& Loeb, A.\ 2004, ApJL, 614, L53
\bibitem[Zhu et al.(2016)]{2016MNRAS.461.1317Z} Zhu, X.-J., Wen, L., Xiong, J., et al.\ 2016, MNRAS, 461, 1317







































\end{thebibliography}
\end{document}